# Drift-diffusion model for spin-polarized transport in a non-degenerate 2DEG controlled by spin-orbit interaction.

Semion Saikin

Center for Quantum Device Technology, Clarkson University, Potsdam, NY 13699, USA

Department of Theoretical Physics, Kazan State University, Kazan 420008, Russia

We apply the Wigner function formalism to derive drift-diffusion transport equations for spin-polarized electrons in a III-V semiconductor single quantum well. Electron spin dynamics is controlled by the linear in momentum spin-orbit interaction. In a studied transport regime an electron momentum scattering rate is appreciably faster than spin dynamics. A set of transport equations is defined in terms of a particle density, spin density, and respective fluxes. The developed model allows studying of coherent dynamics of a non-equilibrium spin polarization. As an example, we consider a stationary transport regime for a heterostructure grown along the (0, 0, 1) crystallographic direction. Due to the interplay of the Rashba and Dresselhaus spin-orbit terms spin dynamics strongly depends on a transport direction. The model is consistent with results of pulse-probe measurement of spin coherence in strained semiconductor layers. It can be useful for studying properties of spin-polarized transport and modeling of spintronic devices operating in the diffusive transport regime.

# Introduction.

Spin-dependent properties of electron transport in semiconductors have recently attracted significant attention of the scientific community in connection with developing field of magneto-electronics or spintronics [1-5]. In comparison with magneto-electronic devices utilizing giant magnetoresistance and tunneling magnetoresistance effects in layered ferromagnetic-metal structures [6], semiconductor spintronic devices promise to be more universal in application due to the ability to adjust a potential variation and spin polarization in an active region of spin devices by external voltages and doping profiles [7-9]. Different designs of transistors and spin-filtering devices utilizing control for the spin polarization in semiconductor structures have been proposed [10-19].

It is in interests of spintronics to build a device that is tolerant to undesirable effects of environment and working at room temperature [1,2,4,5]. However, the crucial phenomena for spintronic devices is a loss of a non-equilibrium spin polarization owing to spin-environment interactions. Functionality of most of proposed devices is sensitive to the temperature, impurities, internal and external fields. Detailed examination of a spin transport problem in semiconductor structures is required for modeling of realistic processes in such devices.

In this work we study spin-polarized transport in a two dimensional non-degenerate electron gas (2DEG) in III-V semiconductor heterostructures in the collision dominated regime [20], where an electron momentum scattering is appreciably faster than spin dynamics. In general, this is applicable for transport in heterostructures with a weak spin-orbit coupling at high temperature. For example, in GaAs/AlGaAs heterostructures for T ~ 100 K or higher an electron momentum scattering is mostly determined by the emission of polar optical phonons [21]. This is valid even at lower temperatures if moderate or strong electric field is applied. The characteristic scattering rate for this process is of the order of several$\cdot 10^{12}$ sec$^{-1}$ [21]. The spin evolution in III-V semiconductors without external magnetic fields is mostly controlled by the spin-orbit interaction [20,22]. It can be characterized by a spin precession frequency, $\Omega$. According to the recent measurements, for a Ga/AlGaAs heterostructure $\Omega \sim 10^{10} - 10^{11}$ sec$^{-1}$ [23,24]. This shows that during one period of a spin precession an electron experiences many collisions. The question is whether this spin transport regime can be useful for spintronics? Since the seminal proposal for the spin field effect transistor (Spin-FET) by Datta and Das [10] utilization of spin-orbit interaction in spintronic devices remains an attractive idea [13-17]. A comprehensive review of the spin-orbit coupling effects for the purposes of spintronics can be found in Ref. [25]. In such devices, due to an electric field dependence of spin-orbit coupling constants, a conventional electric gate can be used to control of coherent electron spin dynamics [10,13-17]. However, the same spin-orbit interaction mechanism leads to spin dephasing due to randomization of an electron momentum (Dyakonov-Perel spin relaxation mechanism [20,22,26]). To avoid undesirable effects of this coupling the device models [10,13-15] was proposed for the ballistic transport regime. In the recent work by Schliemann, Egues, and Loss [16] it was shown that in some cases the spatial electron motion and the spin evolution can be decoupled owing to symmetry of the Rashba [27] and Dresselhaus [28] spin-orbit interaction terms. The effect of momentum scattering on spin coherent dynamics can be diminished and the Spin-FET design [10] is applicable for diffusive transport regime [16,29]. Moreover, the authors have proposed non-ballistic Spin-FET [16] where spin dephasing is controlled by an external gate. Another design of Spin-FET utilizing only non-magnetic materials and operating with a spin dephasing rate in semiconductor heterostructures was proposed in [17]. These devices can be operational in a semiclassical diffusive transport regime that is in interests of this work.

The general drift-diffusion approach for the spin-polarized transport is based on the two, spin-up and spin-down, states model, originally developed for ferromagnetic metals [30,31]. The phenomenological model for non-collinear spin transport including effects of spin-orbit coupling has been developed for the regime where spin dynamics is significantly faster than a momentum scattering rate [32,33]. Though these models can be useful for investigation of a broad class of transport problems in semiconductors [7-9,32-34], they do not include effects of a spin phase memory and are inapplicable for problems where quantum superposition, spin-up-down, states plays an important role [10]. The spin density matrix or spin polarization vector description of a spin state [35-37] is more appropriate for this case.

In our model we use the Wigner function representation for an electron with spin [38]. This approach was utilized before for different transport problems including effects of quantum potential [39], quantum collision [40] and electron transport in magnetic fields [41]. Recently, it has been applied for spin-polarized electron transport in semiconductor heterostructures in the ballistic regime [42]. We consider the semiclassical transport regime where collisions with phonons and impurities control transport properties. We show that in this model the Wigner function transport equation can be reduced to the set of drift-diffusion equations for a particle density, particle current, spin density and spin current.

## Model.

In most of spintronic devices utilizing spin-orbit interaction in semiconductor heterostructures to control spin dynamics [10,13-17], electrons are confined by the effective potential in the direction orthogonal to the semiconductor interface and propagate in the plane of a heterostructure. The effective mass Hamiltonian for an in-plane electron motion in the one subband approximation can be written as

$$H = \frac{\mathbf{p}^2}{2m^*} + V(\mathbf{r}) + H_{SO},$$
$$H_{SO} = \mathbf{pA\sigma}/\hbar.$$
(1)

It is assumed that the electron motion in the direction of quantization can be decoupled from the motion in the plane of a quantum well (QW) and the electron kinetic energy is small in comparison with the subband splitting. The shape of the conduction band is assumed parabolic. Operators of the electron momentum, $\mathbf{p}$, and the spatial position, $\mathbf{r}$, are defined as two dimensional vectors in the plane of QW, while the spin operator, $\boldsymbol{\sigma}$, is a 3-dimensional vector. The potential, $V(\mathbf{r})$, corresponds to the interaction with an electric field oriented in the plane of QW. The spin-orbit interaction term, $H_{SO}$, is written in a general dyadic form linear in an electron momentum. This term is assumed small in comparison with other terms in Hamiltonian, $H$. Matrix elements $A_{j\alpha}$ are constants of spin-orbit interaction, coupling $j$-th component of momentum with $\alpha$-th component of spin. Here, and in the following text, we use Latin letters to index vector or matrix components in spatial dimensions and Greek letters to index components in the spin space. We set the $z$ axis of a spatial coordinate system in the direction orthogonal to the QW plane, while an orientation of a spin coordinate system is left non-specified. An arbitrary rotation of a spin coordinate system will affect the form of the spin-orbit coupling matrix $\mathbf{A}$, but not the general representation of Eq. (1).

A quantum state of an electron with spin can be described by the density matrix operator, $\rho(\mathbf{r},\mathbf{r}',s,s',t)$, which is dependent on two coordinate variables and two spin variables. After the transformation to the new spatial representation,

$$\mathbf{R} = (\mathbf{r}+\mathbf{r}')/2,$$
$$\Delta\mathbf{r} = \mathbf{r}-\mathbf{r}',$$
(2)

the equation for a density matrix will be

$$i\hbar\frac{\partial \rho}{\partial t} = -\frac{\hbar^2}{m^*}\sum_j \frac{\partial^2 \rho}{\partial R_j \partial \Delta r_j} + (V(\mathbf{R}+\Delta\mathbf{r}/2) - V(\mathbf{R}-\Delta\mathbf{r}/2))\rho$$
$$+\frac{i}{2}\sum_{j,\alpha} A_{j\alpha}\left\{\sigma_\alpha, \frac{\partial \rho}{\partial R_j}\right\} + i\sum_{j,\alpha} A_{j\alpha}\left[\sigma_\alpha, \frac{\partial \rho}{\partial \Delta r_j}\right].$$
(3)

The effect of spin-orbit interaction is introduced by the last two terms in Eq. (3), where $[\boldsymbol{\sigma},\ldots]$ and $\{\boldsymbol{\sigma},\ldots\}$ are commutator, and anticommutator with the Pauli spin matrixes respectively. Following the standard transformation to the Wigner function [38],

$$W_{ss'}(\mathbf{R},\mathbf{k},t) = \int \rho(\mathbf{R},\Delta\mathbf{r},s,s',t)e^{-i\mathbf{k}\Delta\mathbf{r}}d^2\Delta r, \qquad (4)$$

and assuming that the potential, $V(\mathbf{r})$, varies slowly and smoothly with the position $\mathbf{r}$, we obtain the transport equation for a single electron with spin

$$\frac{\partial W}{\partial t} + \frac{1}{2}\left\{v_j, \frac{\partial W}{\partial x_j}\right\} - \frac{1}{\hbar}\frac{\partial V}{\partial x_j}\frac{\partial W}{\partial k_j} + ik_j[v_j, W] = \mathrm{St}W. \qquad (5)$$

At the right hand side of Eq. (5) we have included the phenomenological scattering term, St$W$, responsible for interactions of an electron with phonons and non-magnetic impurities. Unlike [35,43] we are interested in a transport regime where electron-electron interaction produces small effects on spin dynamics in comparison with the effects of phonon and impurity scattering [20,22].

In the spin space the velocity operator,

$$v_j = \frac{\partial H}{\partial p_j}, \qquad (6)$$

and the Wigner function, $W$, are (2×2) matrixes, while the potential, $V(\mathbf{r})$, and the electron wave vector $\mathbf{k}$ are scalar variables. The last term on the left hand side of Eq. (5) expresses the spin rotation. Matrix equation (5) can be projected to the set of Pauli matrixes, $\sigma_\alpha$, and unity matrix, $I$, using the following relations for the Wigner function [40]

$$W = \frac{1}{2}(W_n I + W_{\sigma_\alpha}\sigma_\alpha), \qquad (7)$$

and velocity operator

$$v_j = (v_n^j I + v_{\sigma_\alpha}^j \sigma_\alpha),$$

$$v_n^j = \hbar k_j/m^*, \quad v_{\sigma_\alpha}^j = A_{j\alpha}/\hbar. \qquad (8)$$

In the zero order of approximation on the spin-orbit coupling constant, $A_{j\alpha}$, scattering events do not couple different spin components of the Wigner function. The collision term possesses semiclassical form

$$\mathrm{St}W(\mathbf{R},\mathbf{k},t) = \int S(\mathbf{k},\mathbf{k}')(W(\mathbf{R},\mathbf{k}',t) - W(\mathbf{R},\mathbf{k},t))d^2k', \qquad (9)$$

where $S(\mathbf{k},\mathbf{k}')$ is the transition rate for electrons without spin. We use the relaxation time approximation for Eq. (9) with the same set of assumptions, what usually is applied for transport in III-V semiconductors [21]. Corrections to the collision term, St$W$, linear in spin-orbit interaction [44], mix spin polarized components of the Wigner function, $W_\sigma$, with the non-polarized function, $W_n$, and produce the effect of an electron spin polarization by the in-plane electric field [36,44-46]. We do not consider this effect owing to the assumption that spin-orbit coupling is small in comparison with the electron kinetic energy.

To get drift-diffusion transport equations in terms of macroscopic variables we apply the moment expansion procedure [47] to Eq. (5). The (2×2) matrixes of the particle density and current density at the position **R** are defined as

$$n(\mathbf{R}) = \int W d^2 k, \tag{10}$$

and

$$J_j(\mathbf{R}) = \frac{1}{2} \int \{v_j, W\} d^2 k, \tag{11}$$

respectively. Similarly to Eqs. (7) and (8) these matrix variables are projected to the set of the basis matrixes ($\sigma_\alpha$, $\alpha=x,y,z$ and $I$) to get relations for the particle density, spin density, particle current density, and spin current density [35]

$$\begin{aligned} n_n &= \int W_n d^2 k, \\ n_{\sigma_\alpha} &= \int W_{\sigma_\alpha} d^2 k, \end{aligned} \tag{12}$$

$$\begin{aligned} J_n^j &= \int (v_n^j W_n + v_{\sigma_\alpha}^j W_{\sigma_\alpha}) d^2 k, \\ J_{\sigma_\alpha}^j &= \int (v_n^j W_{\sigma_\alpha} + v_{\sigma_\alpha}^j W_n) d^2 k. \end{aligned} \tag{13}$$

The vector of the spin density can be expressed as $n_\sigma(\mathbf{R}) = n_n(\mathbf{R})\mathbf{P}(\mathbf{R})$, where **P(R)** is, normalized to one, spin polarization vector [48] of a small area, $d^2 r$, of 2DEG at the position **R**. The spin density, $n_\sigma(\mathbf{R})$, corresponds to the density of magnetic moment as $\boldsymbol{\mu}(\mathbf{R}) = -g\mu_B n_\sigma(\mathbf{R})$. We assume that the vectors of particle and spin currents, Eq. (13), can be written using the average flow velocity, $\langle v_j \rangle$, as

$$\begin{aligned} J_n^j &= \langle v_j \rangle n_n + \left( v_{\sigma_\alpha}^j - \hbar \langle \Delta k_{\sigma_\alpha}^j \rangle / m^* \right) n_{\sigma_\alpha}, \\ J_{\sigma_\alpha}^j &= \langle v_j \rangle n_{\sigma_\alpha} + \left( v_{\sigma_\alpha}^j - \hbar \langle \Delta k_{\sigma_\alpha}^j \rangle / m^* \right) n_n, \end{aligned} \tag{14}$$

An additional parameter, $\langle \Delta k_{\sigma_\alpha}^j \rangle$, is introduced to define an electron spin-polarized state. For electrons injected from a ferromagnetic contact to QW it is simple to show that $\langle \Delta k_{\sigma_\alpha}^j \rangle = m^* v_{\sigma_\alpha}^j / \hbar$, see Fig. 1(a), owing to the particle density and current density conservation at the interface [49]. Another possible electron spin-polarized state has been considered in the work by Mishschenko and Halperin [42], where $\langle \Delta k^j_{\sigma_\alpha} \rangle = 0$, Fig. 1(b).

In this work we study the first case, where electron spin-polarized state is created with a constant energy rather than a constant electron wave vector. Moreover, we assume that the electron state is inhomogeneously broadened in the momentum space due to temperature effects, and electron velocity can be expanded about the macroscopic flux velocity as

$$v_j = \langle v_j \rangle + \delta v_j. \qquad (15)$$

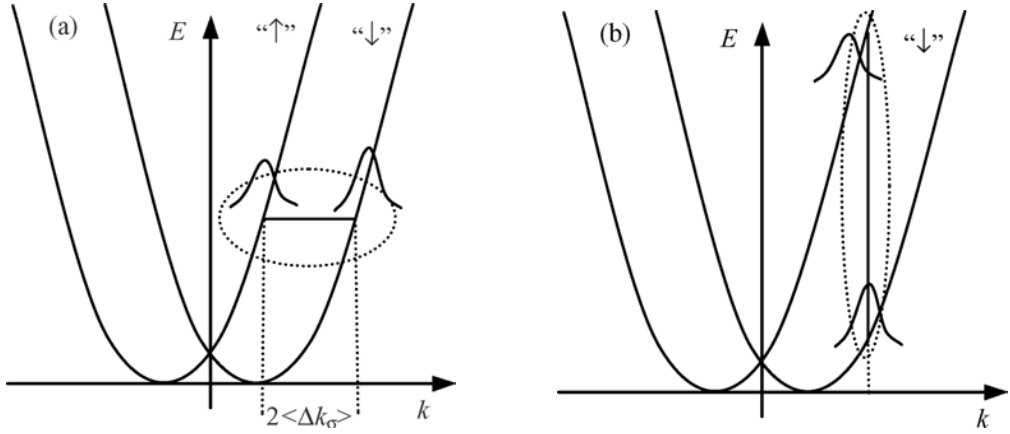

**Figure 1.** Different electron spin-polarized states in 2DEG, (a) electron state created with the energy conservation, (b) electron state created with the wave vector conservation.

The particle and spin conservation equations are obtained by integration of Eq. (5) over an electron wave vector

$$\frac{\partial n_n}{\partial t} + \frac{\partial J^j_n}{\partial x_j} = 0,$$

$$\frac{\partial n_\sigma}{\partial t} + \frac{\partial J^j_\sigma}{\partial x_j} - \frac{2m^*}{\hbar}[v^j_\sigma \times J^j_\sigma] = 0. \qquad (16)$$

The effect of spin-orbit interaction appears in the second equation as the rotational term, where $[a_\sigma \times b_\sigma]$ is used for a vector multiplication in the spin space. This term is proportional to the average flow velocity unlike the case of spin-polarized transport in an external magnetic field [37]. The drift-diffusion equations for the particle and spin currents are derived applying the operator

$$\frac{1}{2}\int \{v^j,...\}d^2k, \qquad (17)$$

to Eq. (5). Assuming the conventional relation, $\langle \delta v_n^j \delta v_n^l \rangle = \delta_{jl} kT/m^*$, we obtain the particle current density and spin current density

$$J_n^j = -\frac{\tau}{m^*}\left(kT\frac{\partial n_n}{\partial x_j} + \frac{\partial V}{\partial x_j}n_n\right),$$

$$J_\sigma^j = -\frac{\tau}{m^*}\left(kT\frac{\partial n_\sigma}{\partial x_j} + \frac{\partial V}{\partial x_j}n_\sigma - \frac{2m^*kT}{\hbar}[v_\sigma^j \times n_\sigma]\right). \quad (18)$$

In Eq. (18) we neglected by terms quadratic in $v_\sigma^j$ mixing polarized and non-polarized components of current. These corrections should be considered in the model, accounting non-conservation of a spin current density, Eq. (11), in systems with a spin-orbit interaction [50]. The term proportional to $[v_\sigma^j \times n_\sigma]$ is responsible for the Dyakonov-Perel spin relaxation [20,22]. Within the applied approximations the equations (16) and (18) do not mix electron transport in different spatial directions. Therefore, the spin-polarized transport in 2DEG can be considered as 1D problem.

The set of the drift-diffusion transport equations (16) and (18) for spin polarized electrons in the presence of the linear in momentum spin-orbit interaction term is the main result of this work. To include effects of an electron-electron interaction in the effective field approximation the transport equations (16) and (18) should be supplemented by the Poisson equation.

## Examples and discussion.

We apply the derived equations to study transport in an asymmetric single QW grown in (0, 0, 1) direction in the crystallographic axes. The electric field in the plane of QW is assumed homogeneous and equal to the external field, $E$. The $x$ axis of the spatial coordinate system is oriented along the electric field and forms angle $\xi$ with the (1, 0, 0) direction in the $xy$ plane. The spin-orbit interaction term, linear in an electron wave vector, is

$$H_{SO} = k_x(\sigma_y(\beta\langle k_z^2\rangle\sin 2\xi - \eta) - \sigma_x\beta\langle k_z^2\rangle\cos 2\xi) +$$
$$k_y(\sigma_x(\beta\langle k_z^2\rangle\sin 2\xi + \eta) + \sigma_y\beta\langle k_z^2\rangle\cos 2\xi), \quad (19)$$

where the spin coordinate system is oriented parallel to the spatial one. Parameters $\eta$ and $\beta$ are Rashba [27] and Dresselhaus [28] spin-orbit coupling constants respectively. For the following derivation we specify a new spin coordinate system. The $z$ spin axis is parallel to the effective magnetic field, produced by spin-orbit interaction, Eq. (19), for an electron propagating along the external electric field. The $y$ spin axis is orthogonal to the QW plane. In this coordinate system the spin-orbit term, $H_{SO}$, Eq. (1), is

$$\begin{aligned}
H_{SO} &= A_{xz}k_x\sigma_z + (A_{yx}\sigma_x + A_{yz}\sigma_z)k_y, \\
A_{xz} &= \sqrt{\eta^2 + (\beta\langle k_z^2\rangle)^2 - 2\eta\beta\langle k_z^2\rangle \sin 2\xi}, \\
A_{yx} &= ((\beta\langle k_z^2\rangle)^2 - \eta^2)/A_{xz}, \\
A_{yz} &= -2\eta\beta\langle k_z^2\rangle \cos 2\xi / A_{xz}.
\end{aligned} \qquad (20)$$

Within the utilized notation the spin polarization of electrons propagating parallel the $x$ axis will precess about the $z$ spin axis. In the case of Rashba spin-orbit interaction only, $\eta \neq 0$, $\beta = 0$, the $z$ spin axis is oriented along the $y$ spatial axis. For non-zero Dresselhaus term, $\eta = 0$, $\beta \neq 0$, it is parallel to the $x$ axis.

The drift-diffusion spin transport equation is

$$\frac{\partial n_\sigma}{\partial t} - \mathbf{D}\frac{\partial^2 n_\sigma}{\partial x^2} - \mathbf{\alpha}\frac{\partial n_\sigma}{\partial x} + \mathbf{\beta} n_\sigma = 0, \qquad (21)$$

where

$$\mathbf{D} = \begin{pmatrix} D & 0 & 0 \\ 0 & D & 0 \\ 0 & 0 & D \end{pmatrix}, \quad \mathbf{\alpha} = \begin{pmatrix} \mu E & 2B_{xz}D & 0 \\ -2B_{xz}D & \mu E & 0 \\ 0 & 0 & \mu E \end{pmatrix},$$

$$\mathbf{\beta} = \begin{pmatrix} D(B_{xz}^2 + B_{yz}^2) & -\mu E B_{xz} & -B_{yx}B_{yz}D \\ \mu E B_{xz} & D(B_{xz}^2 + B_{yx}^2 + B_{yz}^2) & 0 \\ -B_{yx}B_{yz}D & 0 & DB_{yx}^2 \end{pmatrix}, \qquad (22)$$

$$D = \frac{kT\tau}{m^*}, \quad \mu = -\frac{e\tau}{m^*}, \quad B_{j\alpha} = \frac{2m^* A_{j\alpha}}{\hbar^2}.$$

We consider a few examples of spin dynamics in 2DEG using Eq. (21):

**1.** At time $t = 0$ 2DEG is homogeneously polarized, $E = 0$ and only one spin-orbit interaction mechanism (Rashba or Dresselhaus) is responsible for the spin evolution. Eq. (21) is transformed to the spin relaxation equation

$$\frac{\partial n_\sigma}{\partial t} = -\mathbf{\beta} n_\sigma, \qquad (23)$$

where coupling coefficients β are equal to that, what derived in the work [22].

**2.** Stationary injection of spin polarized electrons at the position $x = 0$ into an infinite QW. The spin-orbit constants are coupled by the relation $\eta = \beta \langle k_z^2 \rangle$. Eq. (21) can be diagonalized for any orientation of electron transport with respect to crystallographic directions. The solution is

$$n_{\sigma_x} = n_{\sigma_x}^0 e^{-\left(\frac{\mu E}{2D} + \sqrt{\left(\frac{\mu E}{2D}\right)^2 + B_{yz}^2}\right)x} \cos(B_{xz}x),$$

$$n_{\sigma_y} = n_{\sigma_y}^0 e^{-\left(\frac{\mu E}{2D} + \sqrt{\left(\frac{\mu E}{2D}\right)^2 + B_{yz}^2}\right)x} \sin(B_{xz}x), \qquad (24)$$

$$n_{\sigma_z} = n_{\sigma_z}^0 e^{-\left(\frac{\mu E}{2D} + \left|\frac{\mu E}{2D}\right|\right)x}.$$

Analogously to [16] for the $z$ component of the spin polarization the Dyakonov-Perel spin relaxation mechanism [20,22] is suppressed. The transverse component of spin polarization evolves about the effective magnetic field and decays with the characteristic spin dephasing length, $L_\perp = \left(\frac{\mu E}{2D} + \sqrt{\left(\frac{\mu E}{2D}\right)^2 + B_{yz}^2}\right)^{-1}$. The effect of the electric field, $E$, on the transverse spin dephasing length is similar to that, what was obtained in [51] and later considered in [8,37]. The remarkable property of solution (24) is that the temperature affects coefficients $B_{j\alpha}$ through an effective mass only. Usually, this effect is weak. Therefore, the distribution of the spin polarization should be nearly the same at different temperatures once the relation $\mu E / D$ is conserved.

We estimate the length of the coherent spin precession, $L_p = 2\pi / B_{xz}$, and transverse spin dephasing length, $L_\perp$, for a 10 nm width GaAs/AlGaAs QW. The Dresselhaus spin-orbit constant $\beta = 25.5\,\text{eVÅ}^3$ is taken from the reference [52]. This corresponds to $\beta \langle k_z^2 \rangle = 0.025$ eVÅ. The same order of value for the Rashba spin-orbit coupling constant can be achieved by an appropriate doping of a heterostructure. The calculated values of the spin dephasing and spin precession lengths for different orientations of electron transport are shown in Fig.2.

The essential requirement for the realization of the Spin-FET proposed by Datta and Das [10] is $L_p \ll L_\perp$. As it follows from Fig.2, this relation is valid for transport within the small angle about the (1, -1, 0) direction in crystallographic axes. The applied electric field increases the spin dephasing length while does not affect spin precession length. Moreover, the range of transport directions usable for the Spin-FET [10] is varied with the in-plane electric field.

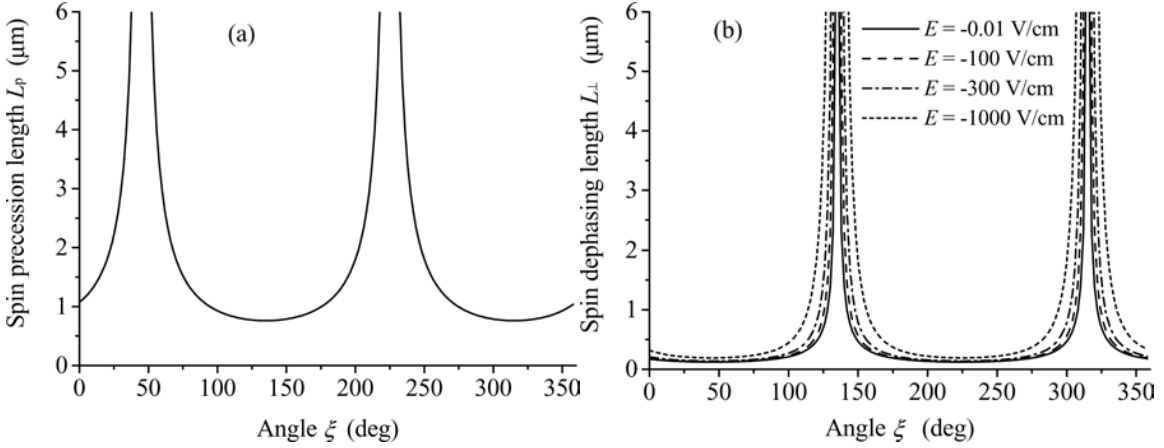

**Figure 2.** Spin precession length (a) and transverse spin dephasing length (b) for different transport orientations with respect to the (0, 0, 1) direction in crystallographic axes at room temperature.

**3.** Stationary injection of spin polarized electrons into a 10 nm width GaAs/AlGaAs QW along the (1, -1, 0) crystallographic direction. At the injection boundary $n_{\sigma_x} = 0$, $n_{\sigma_y} = 0$, $n_{\sigma_z} = n_{\sigma_z}^0$. Spin-orbit coupling constants are not equal. This configuration can be utilized for Spin-FET proposed by Schliemann, Egues, and Loss [16]. The longitudinal spin density component, $n_{\sigma_z}$, decays as

$$n_{\sigma_z} = n_{\sigma_z}^0 e^{-\left(\frac{\mu E}{2D} + \sqrt{\left(\frac{\mu E}{2D}\right)^2 + B_{yx}^2}\right)x}. \tag{25}$$

where $B_{yx} = 2m^*\beta\langle k_z^2\rangle\Delta/\hbar^2$. The calculated spin dephasing length, $L_\parallel$, as a function of the relative difference of the spin-orbit constants, $\Delta = \left|\eta/(\beta\langle k_z^2\rangle) - 1\right|$, is shown in Fig. 3.

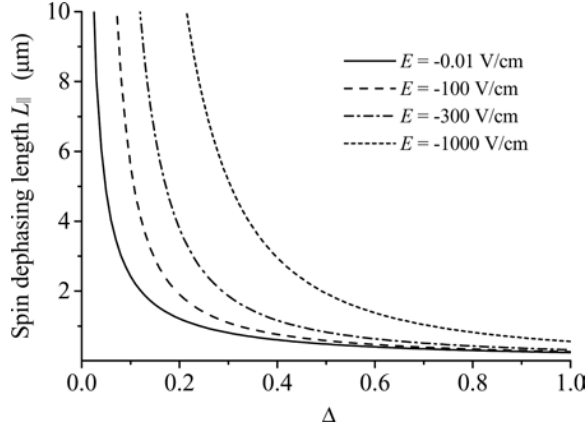

**Figure 3**. Longitudinal spin dephasing length as a function of the relative difference of the spin-orbit constants $\Delta = \left| \eta / \left( \beta \langle k_z^2 \rangle \right) - 1 \right|$ for different values of the applied electric field at T = 300 K. The electric field is along the (1, -1, 0) direction.

Design of the non-ballistic Spin-FET [16] requires efficient modulation of the spin scattering length from $L_\parallel \gg L_D$ to $L_\parallel \ll L_D$ ($L_D$ is a device length) varying the difference between the Rashba and Dresselhaus spin-orbit constants. As follows from our calculations, see Fig. 3, the device operation can be optimized for different relations between $\eta$ and $\beta \langle k_z^2 \rangle$ by the in-plane electric field. For example, if the allowable range of spin dephasing lengths is limited within $L_D/3 < L_\parallel < 3L_D$ then for the device length $L_D = 3\,\mu\text{m}$ $\Delta$ should be varied within $0.03 < \Delta < 0.22$ ($E = -0.01\,\text{V/cm}$), $0.08 < \Delta < 0.3$ ($E = -100\,\text{V/cm}$), $0.13 < \Delta < 0.42$ ($E = -300\,\text{V/cm}$) and $0.22 < \Delta < 0.69$ ($E = -1\,\text{kV/cm}$). Recent experiments demonstrate that 25% variations of parameter $\Delta$ can be easily achieved for GaAs/AlGaAs QWs [23].

Though within the developed model for $\Delta = 0$, the Dyakonov-Perel spin relaxation is completely suppressed for the spatial transport along the (1, -1, 0) direction, other spin scattering mechanisms will determine the spin dephasing. For example, it was shown in [53] that spin-orbit terms cubic in an electron momentum can appreciably modify spin dynamics during the ballistic transport in a quasi-1D structure. The upper limit for the spin dephasing length, possibly, will be defined by the spin scattering at nuclear spins [54]. For high electron concentrations electron-electron collisions will affect spin dynamics even without the momentum dissipation [35,43,55]. Moreover, the relaxation time approximation is a rough model for transport in polar semiconductors [56]. However, we assume that Eqs. (21) and (22) can be valid for more general, field dependent form, of diffusion coefficient and mobility $D(E)$, $\mu(E)$ [21]. These transport parameters can be obtained from a Monte Carlo modeling [57,58].

All-electric measurements of a spin polarization in semiconductor heterostructures [59-61] are complicated by additional spin-independent effects [60]. An existent results do not have single theoretical explanation [32,33,60,62]. However, our model is consent with results obtained in experiments on optical manipulation of spin coherence in strained semiconductor layers [63]. For spin transport along (1, -1, 0) direction the spin polarization rotates for the angle $\varphi = \pi$ on a distance $l \approx 23\,\mu$m. This value is independent of an electric field in the studied range. Moreover, in stronger electric fields the spin coherence is conserved for a longer distance. Both of these features agree with our results, Eq. (24).

The drift-diffusion transport equation, similar to Eq. (21), has been obtained by Pershin [64] within a stochastic approach.

## Conclusions.

We have developed a semiclassical drift-diffusion model for a spin-polarized transport in a non-degenerate 2DEG controlled by the linear in an electron momentum spin-orbit interaction. Effects of an in-plane electric field and transport orientation with respect to crystallographic directions are discussed for a single quantum well grown in the (0, 0, 1) direction in crystallographic axes. The derived model agrees with results of optical measurements of coherent spin dynamics in semiconductor layers. It can be useful for spintronic device modeling.

*Acknowledgements*. I grateful to M.-C. Cheng, D. Mozyrsky, V. Privman and M. Shen for useful discussions. This research was supported by the National Security Agency and Advanced Research and Development Activity under Army Research Office contract DAAD-19-02-1-0035, and by the National Science Foundation, grant DMR-0121146.